\documentclass{aa}
\usepackage{graphicx}
\def\apj{ApJ}
\def\apjs{ApJS}
\def\aap{A\&A}
\def\aas{A\&AS}
\def\aj{AJ}
\def\mn{MNRAS}
\def\pasp{PASP}

\begin{document}

\thesaurus{05(08.05.3, 10.05.1, 10.06.1, 10.07.2)}

\title{Metal-rich globular clusters in the galactic disk: new age
determinations and the relation to halo clusters}

\author{M.~Salaris \and A.~Weiss}

\institute{Max-Planck-Institut f\"ur Astrophysik,
           Karl-Schwarzschild-Str.~1, 85748 Garching,
           Federal Republic of Germany
           }

\offprints{A.~Weiss; e-mail: weiss@mpa-garching.mpg.de } 
\mail{A.~Weiss}

\date{Received; accepted}

\authorrunning{Salaris \& Weiss}
\titlerunning{Galactic Disk Clusters}

\maketitle

\begin{abstract}
New age determinations of the galactic disk globular clusters 47~Tuc,
M71 and NGC6352 have been performed with our up-to-date
$\alpha$-enhanced stellar models. We find that all three clusters are
about 9.2 Gyr old and therefore coeval with the oldest disk white
dwarfs. Several arguments are presented which indicate that the
initial helium content of the stars populating these clusters is close
to the solar one. We also revisit a total of 28 halo clusters, for
which we use an updated [Fe/H] scale.  This new metallicity scale
leads on average to an age reduction of around 0.8 Gyr relative to our
previous results. We compare the predicted cluster distances, which
result from our dating method, with the most recent distances based on
HIPPARCOS parallaxes of local subdwarfs. We further demonstrate that
for the most metal-rich clusters scaled-solar isochrones no longer can
be used to replace $\alpha$-enhanced ones at the same total
metallicity. The implications of the presented age determinations are
discussed in the context of the formation history of the Galaxy.
\end{abstract}

\section{Introduction}
The galactic globular clusters (GC) constitute the fossil record of
the Galaxy formation epoch; their ages therefore 
provide fundamental informations about the timescale  
of the formation process, and put strong constraints on
the age of the universe.

In two recent papers (Salaris, Degl'Innocenti \& Weiss 1997; Salaris
\& Weiss 1997; Papers I \& II) we have redetermined ages for a sample
of GCs by means of new stellar models employing the latest
improvements in stellar input physics, noticeably in equation of state
and opacities. In Paper I we demonstrated that the age of three
extremely metal-poor clusters is around 12 Gyr due to the model
improvements. In Paper II the same models were applied to a sample of
25 halo clusters. We put special emphasis on the careful application
of two distance and reddening-independent methods for determining
absolute and relative ages within groups of clusters of comparable
metallicity. Not only did we confirm the new, low value for the age of
the oldest clusters, but also found a clear correlation between age
and metallicity of the halo clusters. The lowest age in our sample
amounted to only 6.5 Gyr for Ter~7, which, however, might not be one
of the 
typical halo clusters (see the discussion in Paper II), for which the
lowest age was 8 Gyr.  We emphasize that our predicted distances were
in agreement with the first HIPPARCOS-based distance determinations.  
Several
aspects of our results confirmed those of other papers (Chaboyer \&
Kim 1995; Mazzitelli, D'Antona \& Caloi 1995) or have been confirmed
in the following (Sarajedini, Chaboyer \& Demarque 1997).

Since the lowest ages for typical halo clusters were comparable to
those of disk white dwarfs (Salaris et al.\ 1997), it is interesting
to ask how {\sl disk} GC (see, e.g., Armandroff 1989 for a study
about the kinematical properties of this GC system) fit into the
emerging picture of the history of Galaxy formation.  To determine
their ages we have used the same rigorous approach we have been using
successfully for the halo clusters.  We selected three {\sl disk}
clusters, namely 47~Tuc, M71 and NGC6352, with $V$ and $B$ photometry
extending well below the turn-off (TO) point, and with almost the same
metallicity ([Fe/H]=-0.70 for 47~Tuc and M71, [Fe/H]=-0.64 for NGC6352
according to Carretta \& Gratton 1997).  In the case of 47~Tuc and
NGC6352 the absolute age is directly determined by means of the
difference between the Zero Age Horizontal Branch (ZAHB) level and the
TO, the so-called $\triangle V$ age indicator (or {\sl vertical
method}), while for M71, whose photometry shows a poorly populated HB,
the relative age with respect to 47~Tuc is determined by
means of the colour difference between the TO and the base of the Red
Giant Branch (RGB), the so-called $\triangle (B-V)$ age indicator (or
{\sl horizontal method}).

Due to the high metallicity of the {\sl disk} clusters, additional
stellar evolution calculations became necessary. The computational
details as well as the age determination method will be presented
briefly in Sect.~2. The application of our isochrones to the three
clusters will then be discussed in detail in Sect.~3. After that, we
will revisit 
our halo cluster sample, which we extended by another three
objects. Due to new results about the metallicity scale of globular
clusters (Carretta \& Gratton 1997), 
a re-evaluation of their ages became necessary. The updated
results are contained in Sect.~4. In the final section we will discuss
all results.

\section{Computations and age determination method}

For the details of the stellar evolution computations we refer the reader
to Papers I and II. Suffice to repeat here the most important aspects
of the physical input for the evolutionary tracks and isochrones. 

\subsection{Equation of state (EOS)}
We have used the OPAL EOS (Rogers, Swenson \&
Iglesias 1996) supplemented for the regions where the OPAL EOS
is not available, with the EOS described in Paper I.

\subsection{Heavy elements mixture}
For the relative abundances of the elements heavier than helium we
use an $\alpha$-enhanced heavy elements mixture (Table~\ref{t:ae})
selected according to the results presented by Ryan et al.\ (1991)
about chemical abundances in metal-poor field stars. 
%where we have prefered
%values for the most metal-poor objects.
These abundance ratios are also consistent with the results summarized
in Wheeler et al.\ (1989) about the chemical composition of GC stars
and with those by Gratton et al.\ (1986) and Gratton (1987a, b) --
as compiled in Carney (1996) -- about the
$\alpha$-element distribution in the three metal-rich disk clusters
we will study in the next section.
Very recently, McWilliam (1997) has reported $\alpha$-element
enhancements for metal-poor field stars derived from different
authors; these overabundances are again in broad agreement 
with the heavy elements mixture we adopted. Note, however, that the
error-bars in these determinations are so large that a constant
enhancement value for all $\alpha$-elements cannot be excluded
definitely.

Ti is the only element for which our selected enhancement (in agreement
with the Ryan et al.\ (1991) results for the lowest metallicities) does
not agree well with the results of other investigations. Nevertheless,
this is only of minor concern, since Ti 
%(whose abundance contributes for only the 0.02\% to the total metal
%content, and is at least one order of magnitude smaller than in the
%case of the other $\alpha$ elements) 
affects appreciably only the opacities through the TiO molecule, but
in a temperature region that does not affect the evolution of
the stellar models computed for producing our isochrones  
(see, e.g., the discussion on Alexander \& Ferguson 1994).

\begin{table}
\caption{$\alpha$-element enhancements -- expressed in spectroscopic
notation as [element/Fe] -- used
for the computations compared to observed values}
\protect\label{t:ae}
\begin{flushleft}
\begin{tabular}{l|cccc}
Element & (1) & (2) & (3) & (4) \\
\hline
O  & 0.50 & 0.35--0.50 & 0.30--0.50 & 0.31--0.48 \\
Ne & 0.29 &    ---     &    ---     &    ---     \\
Mg & 0.40 & 0.20--0.30 &    0.36    &    ---     \\
Si & 0.30 & 0.25--0.30 &    0.38    & 0.29--0.37 \\
S  & 0.33 &    ---     &    ---     &    ---     \\
Ca & 0.50 & 0.10--0.45 & 0.18--0.47 & 0.09--0.29 \\
Ti & 0.63 & 0.30--0.35 &    0.29    & 0.23--0.40 \\
\hline\noalign{\smallskip}
\noalign{(1): this paper; cf.\ Ryan et al.\ (1991); (2): Wheeler et
al.\ (1989); (3) McWilliam (1997); (4) Carney 1996}
\end{tabular}
\end{flushleft}
\end{table}

\subsection{Opacities}
We have used appropriate and up-to-date opacity tables (taking into
account the effect of molecules at low temperatures) for the
$\alpha$-element enhanced mixtures discussed above for all
temperatures and densities of interest (Rogers \& Iglesias 1992; Alexander \&
Ferguson 1994). Equivalent tables for mixtures with solar metal ratios
were used for scaled-solar ischrones.

\subsection{Mixing length}
A value of 1.80 for the mixing length parameter has been adopted; this
value satisfies the solar constraint and at the same time reproduces
well the RGB of a selected sample of GC in the ($M_{\rm bol},T_{\rm
eff}$) plane, as derived by Frogel et al.\ (1983), once their data are
adjusted to the distance scale given by our ZAHB models.

\subsection{Bolometric corrections and colour transformations}
The conversion of $L/L_\odot$ and $T_{\rm eff}$ into 
($U,\,B,\,V$)-magnitu\-des and colours is performed by adopting
a combination of Buser \& Kurucz (1978 - BK78) and Buser \& Kurucz
(1992 - BK92) transformations (see Paper I). Since in this paper 
we want also to compare the GC distance moduli obtained from our ZAHB
models with the distance
scale set by the HIPPARCOS subdwarfs, we have devoted particular care to
the calibration of the zero point of the bolometric correction scale.

The BK78 and BK92 bolometric corrections (BC) are normalized such that
the maximum value of the BC is zero (see BK78), all others being
negative.  This BC scale has to be calibrated later in such a way as
to reproduce observed V magnitudes of selected stars (as discussed in
BK78 and BK92).  To this purpose we have used the recent empirical BC
determinations by Alonso et al.\ (1995, 1996) for main sequence (MS)
stars of low to solar metallicity. With a constant shift
applied, our BC can reproduce these empirical results for all the
metallicities considered in this paper within $\approx \pm$0.02 mag.
The Alonso et al.\ (1995, 1996) BC
are on a scale where $BC_{\odot}$=-0.12; we therefore adopted $M_{{\rm
bol},\odot}$=4.70 in order to be consistent with the observed solar V
brightness ($M_{V,\odot}$=4.82$\pm$0.02 according to Hayes 1985).

With respect to Paper I and II, where the zero point of our BC was 
calibrated according to the old empirical BC scale for hot MS stars
($T_{\rm eff}> 8000$ K) of nearly solar metallicity by Code et al.\ (1976),
our isochrones in the  $V$-$(B-V)$ plane are now 0.06 mag brighter.
We think that the present recalibration is more reliable since we are using 
new empirical BC for stars with the same metallicity and
the same range of $T_{\rm eff}$ as for GC stars, a fact that  
minimizes possible differential errors in the theoretical
BC varying with metallicity and $T_{\rm eff}$.

Of course, the zero point of the adopted BC scale
does not influence at all the absolute and relative ages obtained by means
of the vertical and horizontal methods used here and in Paper I and
II. However, it is important for the {\sl  a posteriori}\/ calculations
of the distances. 

\subsection{Additional evolutionary calculations}
In addition to the metallicities specified in Paper II we added
consistent calculations for $Z=0.008$ and $Z=0.01$, 
corresponding to ${\rm[Fe/H]}=-0.7$ and $-0.6$ dex. 
The helium content at these
metallicities was either $Y=0.242$, which is identical to that for the
mixture with $Z=0.004$ (the most metal-rich one used in Paper~II), or
respectively $Y=0.254$ and $Y=0.260$ obtained when using $\triangle 
Y/\triangle Z = 3$ as in Paper~II, or $Y=0.273$, which is close to the
initial solar helium abundance as derived from the computation of
solar models (see e.g.\ Schlattl et al.\ 1997). For some $(Y,\,Z)$-combinations
evolutionary calculations for scaled-solar metal abundances were added
for comparison (see Sect.~3.4). With these new calculations our set
of $\alpha$-enhanced isochrones spans the [Fe/H] range
-2.3$\leq$[Fe/H]$\leq$-0.6.

\subsection{Age determination method}
As already mentioned in the introduction,
to determine directly the absolute age of the clusters we use the {\sl vertical
method} (see Stetson et al.\ 1996 for a review and Paper I), which
makes use of the age-dependent brightness difference between TO and
ZAHB. This method is reliable when the cluster photometry shows a
well-populated HB; in this case the ZAHB
level corresponds to the lower envelope of the stellar distribution
along the HB, and it is determined by means of a statistical method 
discussed in Paper II. 
%which, for
%metal-rich clusters generally is fulfilled. As in Paper II, the ZAHB
%level is determined by
%a statistical method identifying the lower envelope of the observed HB
%as the ZAHB level. 
%because RR~Lyr stars are very rare in the clusters
%under consideration.  
When it is not possible to apply the {\sl vertical method},
relative ages between clusters of comparable metallicity are
determined by the {\sl horizontal method}, which uses the
age-dependent colour difference between TO and the base of the RGB
(see VandenBerg et al.\ 1990).

After the age has been determined, we check our results with various
tests: (i) we verify that the corresponding isochrone fits the cluster
CMD sufficiently well; (ii) we compare our predicted reddening
(obtained from fitting the unevolved isochrone main sequence to the observed one)
with literature values; (iii) we use the theoretical ZAHB brightness
to derive the distance modulus of the cluster, and compare this with
independent determinations.

\section{Metal-rich disk clusters}

\subsection{NGC104 (47~Tuc)}

The most prominent disk cluster is certainly 47~Tuc on the southern 
hemisphere. We used the data from the paper by
Hesser et al.~(1987). The metallicity of 47~Tuc has been determined as
${\rm [Fe/H]} = -0.70\pm0.10$ by Carretta \& Gratton (1997) and is
confirmed by Cannon et al.~(1997). 
Using the vertical method for deriving the cluster age,
Hesser et al.\ (1987) determined t=13.5$\pm$2.0 Gyr for 47~Tuc from
oxygen-enhanced isochrones, 
while Chaboyer \& Kim (1995) obtained, by applying their
iso\-chrones without diffusion (computed by using the OPAL equation of state)
to the data by Hesser et al.\ (1987),
t=13.23$\pm$1.7 Gyr; Mazzitelli et al.\ (1995) derived ages of 12--14 Gyr
from the same observational data.
Very recently Gratton et al.\ (1997) determined the age of 47~Tuc 
matching the observed TO position (from Hesser et al.\ 1987) with 
several sets of theoretical isochrones, after deriving the distance to
the cluster by means of the main sequence (MS) fitting using HIPPARCOS sub\-dwarfs;
they got ages in the range 9.6-10.8 Gyr, with a random error of 12\% on the 
single age values.

There have been several suggestions that the helium content of 47~Tuc
stars is close to the solar value. Alonso et al.\ (1997) find that
$\triangle V_{\rm TO}^{\rm HB}$ is almost the same in NGC6366 (a
metal-rich halo cluster, with a metallicity close to that of 47
Tuc, see Table~\ref{t:hc}) and 47~Tuc, implying the same age for both
clusters; however, $\triangle(B-V)$ is definitely larger for 47~Tuc,
thus indicating that this cluster has to be younger than NGC6366.
This means that two age indicators give contradicting results.  Since
$\triangle(B-V)$ is largely independent of the helium content while
$\triangle V_{\rm TO}^{\rm HB}$ is dependent (as shown by Salaris et
al.\ 1994 and confirmed by our computations), they take this as
evidence for an enriched helium content in 47~Tuc with respect to
NGC6366, concluding that a value Y=0.27-0.28 for 47~Tuc could explain
this discrepancy.  Buzzoni et al.\ (1983) and Hesser et al.\ (1987)
found that the $Y$-dependent R-parameter, relating the number of stars
on the HB and on the RGB at magnitudes brighter than the HB, is
$1.75\pm0.21$, respectively  $1.86\pm0.36$, corresponding to a
helium content of $0.27\pm0.02$ and \ $0.28\pm0.04$ according to the
calibration by Buzzoni et al.\ (1983).  

We investigated independently
another helium abundance indicator: the brightness difference between
a point on the lower, unevolved MS, that is, for $M_{V}\ge 6.0$,
and the ZAHB at the same colour, $\triangle V_{\rm MS}^{\rm ZAHB}$ (a
slightly different definition of this He indicator was already given
by Caputo et al.\ 1983). Since the MS is shifted at lower luminosities
and the ZAHB at higher ones for an increase of the initial helium abundance
in stellar models, the difference $\triangle V_{\rm MS}^{\rm ZAHB}$
results to be a good
indicator of the stellar initial helium content, that complements the R-parameter.

From our definition of $\triangle V_{\rm MS}^{\rm ZAHB}$ it turns out that 
this indicator can only be used in
clusters with deep MS photometry, an HB redder than the TO and a
well-defined reddening to allow for a precise identification of the
corresponding point on the theoretical isochrone; all three conditions
are fulfilled for 47~Tuc ($E(B-V)=0.04\pm0.02$; Zinn 1980). Note that
this brightness difference is to very good approximation
age-independent and neither influenced by the mixing-length parameter
as we have verified from our isochrones.  At $(B-V)_{0}=0.76$ our
models provide, for [Fe/H] values around [Fe/H]$= -0.7$: $Y = -0.319 +
0.13 \cdot \triangle V_{\rm MS}^{\rm ZAHB} + 0.21 \cdot {\rm [Fe/H]}$.
Taking the reddening (and the associated error) from Zinn (1980), we
derived $\triangle V_{\rm MS}^{\rm ZAHB}=5.66\pm0.12$ and obtained,
considering also the error of $\pm 0.10$ dex in [Fe/H],
$Y=0.270\pm0.026$, in excellent agreement with the quoted values from
the R-parameter.  Consequently, we have used isochrones with solar
helium content to determine the age of 47~Tuc.

Due to the very well-populated red HB, the identification of the ZAHB
level is very easy and unambiguous; applying the method devised in
Paper II we derive $V_{\rm ZAHB}=14.12\pm0.04$.
%(that provides a
%distance modulus $(m-M)_{V}$=13.48).  
The TO-brightness is more difficult to determine, since (as already
noted by Grundahl 1996) not many stars populate this region of the
colour-magnitude diagram provided by Hesser et al.\ (1987).  Hesser et
al.~(1987) give a value of $V_{\rm TO} = 17.70$, although the central
$V$-value of the bluest brightness bin of their main line is at
17.65. Renzini \& Fusi Pecci (1987) quote $V_{\rm TO}$ = 17.65, while
Buonanno et al.~(1997), using the same raw data, determine it to be at
17.50.  We have redetermined the TO from the published photometric
data of Hesser et al.\ (1987). Fitting a parabola to the stars in the
TO region, we obtain $V_{\rm TO} = 17.65\pm0.10$, in good agreement
with the results by Renzini \& Fusi Pecci (1987) and Hesser et al.\
(1987).  We also used two alternative points along the main line for
the absolute age determination, as recently suggested in the
literature: both points are 0.05 mag redder than the (well-determined)
colour of the TO, but either on the MS (Buonanno et al.\ 1997) or on
the subgiant branch (Chaboyer et al. 1996). It turns out that the ages
as determined from the visual brightness difference between ZAHB and
these two points are in mutual agreement (respectively 8.8 and 9.0
Gyr) and agree well with the value of $9.2\pm1.0$ obtained from
$V_{\rm TO}$ = 17.65$\pm$0.10. This age is considerably lower than the
values quoted at the beginning of this section, but comparible to that
based on HIPPARCOS-parallaxes (Gratton et al.\ 1997).

\begin{figure}
\includegraphics[draft=false,scale=0.40]{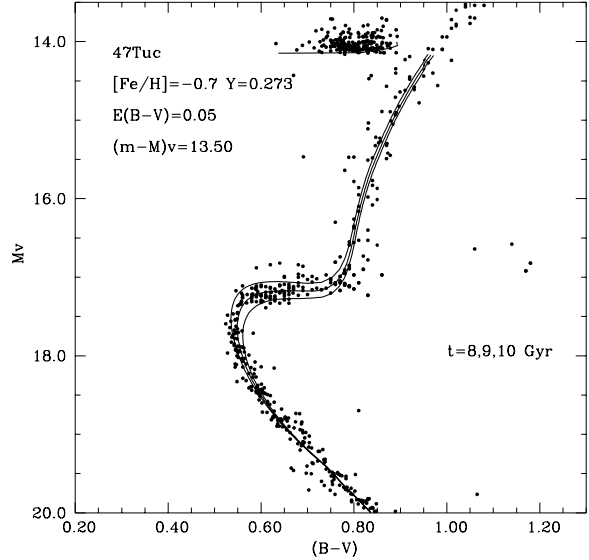}
\caption[]{Isochrones of 8-10 Gyr for the composition $(Y,\,Z)=(0.273,\,0.008)$
($\alpha$-enhanced) applied to 47~Tuc. The data are from Hesser et
al.~(1987)}
\protect\label{f:tuch}
\end{figure}

In Fig.~\ref{f:tuch} we show the fit of representative isochrones for
the mixture $(Y,\,Z)=(0.273,\,0.008)$ ($\alpha$-enhanced). The
agreement with the observations is very good; the reddening we get is
E(B-V)=0.05, consistent with the value quoted by Zinn (1980), which we
have used for the He determination from $\triangle V_{\rm MS}^{\rm
ZAHB}$. The derived apparent distance modulus is $(m-M)_{V}$=13.50.

\begin{figure}
\includegraphics[draft=false,scale=0.40]{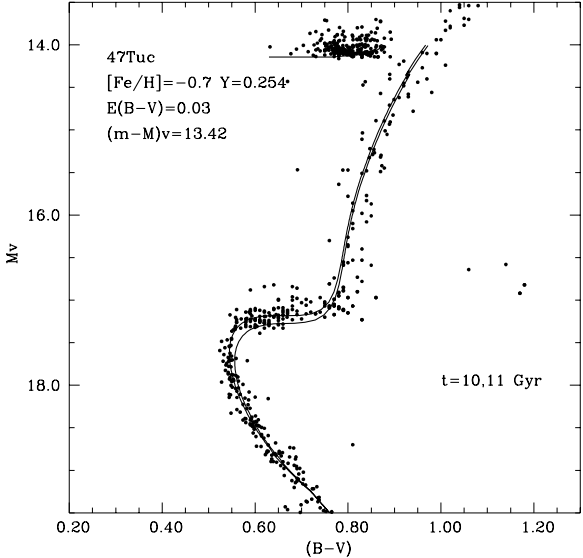}
\caption[]{Isochrones of 10-11 Gyr for the composition $(Y,\,Z)=(0.254,\,0.008)$
($\alpha$-enhanced) applied to 47~Tuc. The observational data are the
same as in Fig.~1}
\protect\label{f:tucl}
\end{figure}

For comparison, Fig.~\ref{f:tucl} shows the fit of 
representative isochrones for
the mixture $(Y,\,Z)=(0.254,\,0.008)$ ($\alpha$-enhanced)
corresponding to our standard case $\triangle Y/\triangle Z = 3$.
Note that the agreement between theoretical and observed RGB is
slightly worse;
from the vertical method one gets an age higher by 1.1 Gyr with
respect to the age obtained adopting $Y=0.273$. We consider this age
as a possible upper limit, since $Y=0.254$ is close to the lower limit
of the possible range of the original He content as derived by us.

\subsection{NGC6838 (M71)}

M71 is a disk cluster about 3.5 kpc from the Sun. Its metallicity is,
according to Carretta \& Gratton (1997), ${\rm [Fe/H]} = -0.70\pm0.10$.
%Carney (1996) gives $\rm{[O/Fe]}=0.4$ and 0.3 for other
%$\alpha$-elements. This is very close to our assumption about
%$\alpha$-enrichment in the models. 
We use the photometric data and
mean line from Hodder et al.\ (1992). 
%the TO-brightness being $V_{\rm
%TO}=18.04\pm0.10$.  
Due to the sparsely populated HB, the application
of the vertical method is not straightforward. We therefore determined
the age of M71 by means of the horizontal method.  If the mean line is
registered to that of 47~Tuc in the same way as in Paper~II (VandenBerg
et al.\ 1990), this gives the same age as for 47~Tuc (as already noted
by Richer et al.\ 1996, Alonso et al.\ 1997), $9.2\pm1.2$ Gyr (in the
error we considered also the contribution due to the
formal uncertainty in determining the relative position of the two RGB
as discussed in Paper II).  
Since the horizontal method is very weakly affected by differences in
the helium content among the clusters (see previous discussion about 47~Tuc),
the relative age between 47~Tuc and M71 that we get is therefore
independent of the possible difference in helium abundance between
these two clusters.
Buonanno \& Iannicola (1995) have determined the helium content to be
$Y=0.29\pm0.03$ from the R-parameter, consistent with the helium
content of 47~Tuc. 

\begin{figure}
\includegraphics[draft=false,scale=0.40]{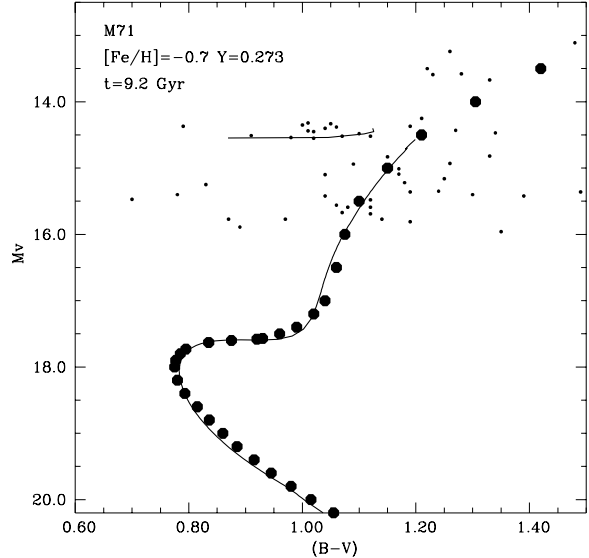}
\caption[]{Isochrone of 9.2 Gyr for the composition $(Y,\,Z)=(0.273,\,0.008)$
($\alpha$-enhanced) applied to M71. The observational data and the
mean line are from Hodder et al.\ (1992). For the cluster MS and
subgiant branch only the main line is displayed.}
\protect\label{f:m71}
\end{figure}

In Fig.~\ref{f:m71} we show our ZAHB and isochrone (Z=0.008 and
Y=0.273) for t=9.2 Gyr superimposed to the colour-magnitude diagram of
M71 in such a way that the theoretical TO matches the position of the
observed one. The predicted ZAHB-level from the isochrone agrees with
the lower envelope of the sparsely populated HB. The vertical method
-- if applicable -- therefore would result in a similar age as the
relative age determination by the horizontal method. The reddening as
obtained from the isochrone is 0.28 and compares well with the value
0.27$\pm$0.03 derived by Zinn (1980).

\subsection{NGC6352}

NGC6352 has been observed lately by Buonanno et al.\ (in preparation);
TO- and ZAHB-brightness of its red HB $V_{\rm TO}$ and $V_{\rm ZAHB}$ 
are, however,
already given in Buonanno et al.\ (1997); they are, respectively, $18.73\pm0.06$
and $15.20\pm0.06$. The value of the horizontal age parameter is also
given, $\triangle(B-V)=0.285\pm0.010$. 

The iron content of NGC6352 given by Carretta \& Gratton (1997) is
[Fe/H]=$-0.64\pm0.10$, consistent within the error bar with the
metallicity of 47~Tuc and M71.  Using the same isochrones applied to
47 Tuc (Z=0.008 and Y=0.273) we obtain from the vertical method an age
of t=$9.4\pm0.6$ Gyr.  For verifying the validity of our assumption
that the helium content of NGC6352 is the same as that of 47~Tuc, we
can use the horizontal method for determining the NGC6352 age relative
to 47~Tuc; we get an age difference by $0.4$ Gyr, NGC6352 being older,
completely consistent with the age from the vertical method.

\subsection{Scaled-solar vs.\ $\alpha$-enhanced isochrones}

\begin{figure}
\includegraphics[scale=0.40,draft=false]{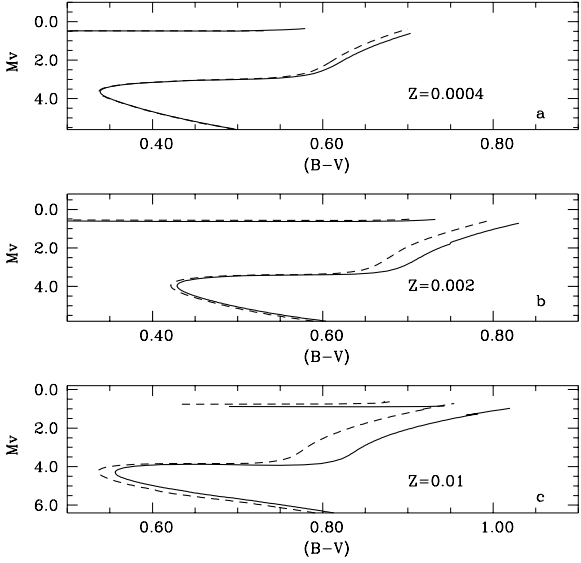}
\caption[]{Comparison of $\alpha$-enhanced (dashed line) and
scaled-solar (solid line) isochrones (t=10 Gyr) and ZAHB with total
metallicity Z=0.0004 (panel a) Z=0.002 (b) and Z=0.01 (c).}
\protect\label{f:ae}
\end{figure}

Since Salaris et al.\ (1993) it has been known that at least for the
lowest metallicities the effect of $\alpha$-element enhancement on the
evolutionary tracks and isochrones can be simulated by using a 
scaled-solar mixture of total metallicity equal to the actual one, provided
that [C+N+O+Ne/ Mg+Si+S+Ca+Fe] $\approx 0$ in the $\alpha$-enhanced
mixture (this condition is automatically fulfilled, adopting current
determinations of the solar heavy elements mixture, if all
$\alpha$-elements are enhanced by the same factor).  Although this
result was based on calculations without low-temperature opacity
tables with $\alpha$-enhancement, it was later confirmed by us in
Paper~I, and it was again verified by Salaris \& Cassisi (1996)
employing low-temperature $\alpha$-enhanced molecular opacities.  On
the other hand, Weiss et al.\ (1994) found that for solar or
super-solar total metallicity, the element abundance ratios within the
metals {\em do} influence the evolution. In these calculations, again
only high-temperat\-ure ($kT > 1$~eV) $\alpha$-enhanced opacity tables
were available. Obviously, there exists a value for the metallicity,
above which $\alpha$-element enhancement -- if present -- has to be
taken into account, not just because it raises the total metallicity,
but also because the distribution of elements influences the evolution
significantly. Because the disk clusters have a metallicity of
$Z\approx Z_\odot/2$, this is of interest for the age
determination. (Recall that for the results of this section, 
$\alpha$-enhancement, also in the opacities for all
temperatures, {\em is} taken into account.)

In Fig.~\ref{f:ae} (panel a) we compare isochrones (for t=10 Gyr) and
ZAHB with a total metallicity Z=0.0004. Solid lines are for the case
that all metals are in scaled-solar abundance ratios, the dotted ones
for the case of $\alpha$-enhance\-ment.  Even at this low metallicity,
the RGB colours differ slightly, a fact that seems to contradict the
claim made by Salaris et al.\ (1993).  However, one has to recall that
their conclusions applied to the case when [C+N+O+Ne/Mg+ Si+S+Ca+Fe]
$\approx$0, while in our heavy elements mixture (see Sect.~2.2) it is
$\approx$ 0.12.  The relatively larger enhancement of oxygen with
respect to the average enhancement of all other $\alpha$-elements
explains the small colour differences along the RGB. The ZAHB and TO
location of the $\alpha$-enhanced isochrone agree perfectly with the
scaled-solar one, therefore the same ages would be derived at this
total metallicity when using scaled-solar or $\alpha$-enhanced
models. Also the relative ages obtained from the horizontal method are
in close agreement, because the small difference in the absolute
colour of the RGB does not affect the $\Delta(B-V)$ scaling
with age.  At Z=0.002 (panel b) small colour differences along the MS
appear; TO and ZAHB luminosities are slightly lower for the
scaled-solar isochrones, but the brightness difference between TO and
ZAHB is virtually unchanged.  The colour differences along the RGB are
larger than in the previous case.  

The case of Z=0.01 (corresponding to [Fe/H]=-0.6 for the
$\alpha$-enhanced isochrones), a metallicity close to the one of the
disk GC considered in our work, is shown in panel c of the same
figure.  At this metallicity the two sets of isochrones appear to be
very different: the scaled-solar ZAHB is dimmer than the corresponding
$\alpha$-enhanced one and the scaled-solar isochrone is redder and its
TO brightness lower. The latter behaviour is in agreement with the
result of Weiss et al.\ (1994) obtained from a comparison of
evolutionary tracks\footnote{The line types for these tracks have been
exchanged erroneously in the caption of their Fig.~4, as is obvious
from comparing the tracks with those in their Fig.~3}.  At this
metallicity, the use of scaled-solar isochrones for deriving the age
of GC using the vertical method leads to ages higher by $\approx$1.0
Gyr with respect to ages determined by means of the appropriate
$\alpha$-enhanced isochrones.

An further point has to be mentioned before concluding this
section: when using the MS fitting technique for deriving the
distance of a GC,
one has  to correct the observed local subdwarf colours to
the colours they would have 
at the metallicity of the cluster. 
To do so one needs the derivative $\triangle (B-V)/\triangle Z $
(at fixed $V$ magnitudes), which is taken from
theoretical isochrones. At metallicities $Z\ge 0.002$, the use of scaled-solar
isochrones can introduce a systematic error in the derived distances. Considering
for example $M_{V}$=6.0, the $(B-V)$ colours at $Z=0.01$ and $Z=0.002$
differ by 0.104 mag, while this difference is equal to
0.116 mag in the case of scaled-solar isochrones. Since the slope of
the MS  $\triangle M_{V}/\triangle (B-V)$ is $\approx 5$ at this
magnitude, the use of scaled-solar isochrones in place of
$\alpha$-enhanced ones would give
a distance modulus too high by $\approx$0.06 mag, if the subdwarfs
have the lower, but the cluster stars the higher metallicity.

\section{An update of the halo cluster sample}

\begin{table*}
\caption{Age determinations for a sample of 28 halo clusters and the
three disk clusters of the present paper. Column 2 contains the iron
abundance as given in Carretta \& Gratton (1997), while column 3 gives
the total metallicity (including $\alpha$-element enhancement). Next,
the age in Gyr as obtained from our analysis follows in column
4. Distance modulus and reddening are listed in the remaining
columns. Note that only for the reference clusters these are obtained
directly. For all other clusters the reddenings are
determined indirectly (see text). In this case, the numbers are in
brackets}
\protect\label{t:hc}
\begin{flushleft}
\begin{tabular}{lrrrrr}
\hline\noalign{\smallskip} 
Cluster &$\rm [Fe/H]$& $Z$ & Age & $(m-M)_V$ & $E(B-V)$  \\ 
\noalign{\smallskip} \hline\noalign{\smallskip} 
{\bf Halo Clusters} &  \multicolumn{3}{c}{$-2.15\leq{\rm[Fe/H]}<-1.75$} \\
 NGC4590~(M68)& -1.99 & $4.2\cdot10^{-4}$ & 11.4$\pm$1.0 &  $15.26$ &
$0.05$ \\
 NGC7078~(M15)& -2.12 & $3.0\cdot10^{-4}$ & 11.4$\pm$1.1 & & $(0.11)$ \\
 NGC6341~(M92)& -2.16 & $2.9\cdot10^{-4}$ & 11.0$\pm$1.1 & & $(0.04)$ \\
 NGC7099~(M30)& -1.91 & $5.4\cdot10^{-4}$ & 11.9$\pm$1.1 & & $(0.04)$ \\
 NGC6397      & -1.82 & $6.2\cdot10^{-4}$ & 11.4$\pm$1.1 & & $(0.18)$ \\
 NGC2298      & -1.74 & $7.4\cdot10^{-4}$ & 11.7$\pm$1.1 & & $(0.22)$ \\
 Arp2         & -1.76 & $7.1\cdot10^{-4}$ & ~9.7$\pm$1.1 & & $(0.13)$ \\
 Rup106       & -1.78 & $6.8\cdot10^{-4}$ & ~9.1$\pm$1.1 & & $(0.22)$ \\
& \multicolumn{3}{c}{$-1.75\leq{\rm[Fe/H]}<-1.30$} \\
 NGC6584      & -1.30 & $2.0\cdot10^{-3}$ & 10.1$\pm$1.0 &  $16.05$ &
$0.10$ \\
 NGC5272~(M3) & -1.34 & $1.9\cdot10^{-3}$ & 10.1$\pm$1.1 & & $(0.01)$ \\
 NGC1904~(M79)& -1.37 & $1.7\cdot10^{-3}$ & 10.1$\pm$1.1 & & $(0.03)$ \\
 NGC6752      & -1.43 & $1.5\cdot10^{-3}$ & ~9.6$\pm$1.1 & & $(0.03)$ \\
 NGC6254~(M10)& -1.41 & $1.8\cdot10^{-3}$ & 10.1$\pm$1.1 & & $(0.28)$ \\
 NGC7492      & -1.61 & $1.0\cdot10^{-3}$ & 10.1$\pm$1.1 & & $(0.03)$ \\
 NGC6101      & -1.60 & $1.0\cdot10^{-3}$ & 10.9$\pm$1.1 & & $(0.09)$ \\
 NGC5897      & -1.59 & $1.0\cdot10^{-3}$ & 10.1$\pm$1.1 & & $(0.10)$ \\
& \multicolumn{3}{c}{$-1.30\leq{\rm[Fe/H]}<-0.90$} \\
 NGC5904~(M5) & -1.11 & $3.2\cdot10^{-3}$ &  9.9$\pm$0.7 &  $14.53$ &
$0.02$ \\
 NGC3201      & -1.23 & $2.4\cdot10^{-3}$ &  9.9$\pm$0.8 &  &
$(0.24)$ \\
 Pal5         & -1.24 & $2.4\cdot10^{-3}$ & ~8.3$\pm$0.8 & & $(0.05)$ \\
 NGC288       & -1.07 & $3.6\cdot10^{-3}$ & ~8.8$\pm$0.8 & & $(0.02)$ \\
 NGC362       & -1.15 & $3.1\cdot10^{-3}$ & ~8.7$\pm$0.8 & & $(0.04)$ \\
 NGC1851      & -1.14 & $3.0\cdot10^{-3}$ & ~7.9$\pm$0.8 & & $(0.08)$ \\
 Pal12        & -1.00 & $4.1\cdot10^{-3}$ & ~6.6$\pm$0.8 & & $(0.02)$ \\
 NGC1261      & -1.09 & $3.3\cdot10^{-3}$ & 10.9$\pm$0.8 & & $(0.00)$ \\
& \multicolumn{3}{c}{$-0.90\leq{\rm[Fe/H]}<-0.8$}\\
NGC6171~(M107)& -0.87 & $5.5\cdot10^{-3}$ & 10.4$\pm$1.0 &  $15.04$ &
$0.36$ \\
  NGC6652      & -0.81 & $6.3\cdot10^{-3}$ & ~8.0$\pm$1.1 &  $15.26$ &
$0.22$ \\
 Ter7         & -0.87 & $5.5\cdot10^{-3}$ & ~7.1$\pm$1.1 & & $(0.07)$ \\
 NGC6366      & -0.87 & $5.5\cdot10^{-3}$ & 12.2$\pm$1.1 & & $(0.72)$ \\
 \noalign{\smallskip} \hline
{\bf Disk Clusters} & \multicolumn{3}{c}{$-0.9\leq{\rm[Fe/H]}<-0.6$} \\
NGC104~(47~Tuc)& -0.70 & $8.1\cdot10^{-3}$ & $9.2\pm1.0$ & $13.50$ & $0.05$ \\
NGC6838~(M71) & -0.70 & $8.1\cdot10^{-3}$ & $9.2\pm1.1$ &  & $(0.28)$ \\
NGC6352       & -0.64 & $9.3\cdot10^{-3}$ & $9.4\pm0.6$ & $14.56$ & --- \\
 \noalign{\smallskip} \hline
& \multicolumn{3}{c}{Scaled-solar metallicity clusters:}\\
Rup106 & -1.78 & $3.0\cdot10^{-4}$ & $9.9\pm1.1$ & & $(0.24)$\\
Pal12  & -1.00 & $1.8\cdot10^{-3}$ & $7.7\pm0.8$ & & $(0.05)$\\
\noalign{\smallskip} \hline
\end{tabular}
\end{flushleft}
\end{table*}

\subsection{Influence of a new metallicity scale}

Carretta \& Gratton (1997) recently have redetermined the metallicity
of a number of globular clusters and on this basis recalibrated the
Zinn \& West (1984) metallicity scale. As a consequence, the
metallicities of the 25 clusters investigated in Paper~II increased
by on average $\approx 0.20$ dex. This effect has not been taken into
account in Paper~II, but it was estimated that ages should be affected
by less than the typical errors in our ages ($\approx \pm1.1$
Gyr as 1--2 $\sigma$ error). For completeness, however, 
we repeated the analysis of Paper~II for the present paper. 
The new results are shown in
Table~\ref{t:hc}. 
In the case of Arp2 and Rup106 we determined the metallicity reported
in Table~\ref{t:hc} recalibrating on the Carretta \& Gratton (1997) scale
the RGB morphological metallicity indicators used for determining the
metallicities given in Paper~II.

In addition to the 25 clusters of Paper~II, we have
added three other halo clusters: 
NGC5897 (Sarajedini 1992) and NGC6101 (Sarajedini \& Da
Costa 1991), both
intermediate metal-poor clusters, and NGC1261 (Bolte \& Marleau 1989), an
intermediate metal-rich cluster. Note that the cluster NGC3201 moved
one metallicity group upwards. Interestingly, since this cluster
allows the application of the vertical method, we used it for
checking the consistency of ages in the intermediate metal-poor group
in Paper~II, while we can do the same test now for the intermediate
metal-rich group. In both cases, the relative age of NGC3201 with
respect to the template cluster in the same metallicity group 
as derived from the horizontal method is 
consistent with its absolute age as derived from
the vertical method.
On average, absolute cluster ages are,
compared to Paper~I, reduced by $\approx 0.8$ Gyr due to the
increased metallicity. 

In Table~\ref{t:hc}, last two columns, we have listed 
direct determinations of distance modulus and reddening for the
reference clusters (see Sect.~2.7), and indicative reddenings for all
others. The latter have been estimated
by means of the following procedure: once the age has
been determined by the horizontal method, the absolute colour of
the TO is known from the theoretical isochrones. Comparison with the
observed TO then gives the reddening (the errors associated with this
procedure, due to the error on the cluster ages, are of the order of
$\pm$0.02 mag). From the theoretical TO one could also obtain a
distance by comparing it with the observed apparent
TO-magnitude. Since this, however, is very uncertain in some cases
(see the discussion about M15 in Paper~I), we did not list them. The
reader may derive estimates for the distances by employing our Eq.~1
(see below).

There has been some concern that a few clusters -- Rup106 and Pal12 in
our sample -- do not show $\alpha$-element enrichment (Brown et al.\
1997). Although these are preliminary results based on only two stars
in each cluster, and still have to be confirmed, we have
investigated by how much the ages would be changed, mainly due to the
reduced total metallicity. In Table~\ref{t:hc} we have therefore added
ages for both clusters based on this assumption.  To obtain them, we
determined the absolute age of the reference cluster in their group
(namely, M68 and M5) from the vertical method by means of
appropriately scaled-solar isochrones (thus assuming for the reference
clusters [$\alpha$/Fe]=0, too) and then we derived the differential
age of Rup106 and Pal12 with respect to M68 and M5 by means of the
horizontal method (using the same set of scaled-solar isochrones). The
net effect amounts to an age increase by, respectively, 0.8 Gyr and
1.1 Gyr. We note that the scaled-solar isochrone
used for Pal12 fits better to the observed RGB than
the one including $\alpha$-element
enhancement. This might be taken as independent, theoretical support
for the claim that this cluster does not show enriched
$\alpha$-elements.

\subsection{Predicted vs.\ HIPPARCOS-based distances}

As emphasized already in Papers I and II, our method to determine
cluster ages is independent of any distance determination. Rather,
the vertical method, which we use for the absolute age of the
reference clusters, together with the theoretical models, allows to predict a
cluster distance. This predicted value can then be compared with other
distance determinations to check for the reliability of our models.

In Paper~I, we have shown that we are in agreement with some of the
relations relating absolute RR~Lyrae brightness with metallicity. In
Paper~II, this point was further emphasized, but since the very first
HIPPARCOS-paralla\-xes had become available, we also compared with
results relevant for cluster distances. Again, we found that we were
consistent with these results, though some of them were in conflict
with each other.

Since then, a number of papers have appeared, which are devoted to the
question of GC distances obtained from the fitting of their MS to
subdwarfs, whose distances can be determined by HIP\-PARCOS-parallaxes
(Reid 1997, 1998; Grat\-ton et al.~1997; Chaboyer et al.~1997; Pont et
al.~1998). In these papers, different samples of subdwarfs (sometimes
including also objects with only ground-based parallaxes), different bias
corrections, different metallicities for the same subdwarfs or
globular clusters have been used, and the results do not agree in all
cases (e.g.\ the case of M92: Reid 1997, Pont et al.~1998 and Gratton
et al.~1997).

For our reference clusters (see Table~\ref{t:hc}) we found
HIP\-PAR\-COS-parallaxes based distance determinations (in the
following for simplicity called ``HIP\-PARCOS-distances'') for M68
(Reid 1997; Gratton et al.~1997), M5 (Reid 1997; Gratton et al.~1997;
Chaboyer et al.~1997), and 47~Tuc (Gratton et al.~1997, Reid 1998).  
Moreover, we
can also compare our theoretical $M_{V}$(ZAHB)-[Fe/H] relation (taken
at the average temperature of the RR Lyrae instability strip, $\log
T_{\rm eff}$$\approx$3.85) with the observational one derived by
Gratton et al.\ (1997) for their sample of clusters by means of
MS fitting.

\begin{table}
\caption{Predicted and HIPPARCOS distances for reference clusters
(Table~2), given as the apparent distance modulus $(m-M)_V$. See
text for more explanations}
\protect\label{t:hd}
\begin{flushleft}
\begin{tabular}{lrrl}
Name & this paper & Gratton et al.~1997 & other ref. \\
\hline\noalign{\smallskip}
M68 & $15.26\pm0.05$ & $15.31\pm0.08$ & $15.35\pm0.10^1$\\
M5  & $14.53\pm0.05$ & $14.60\pm0.07$ & $14.54\pm0.10^1$\\
    &                &                & $14.51\pm0.09^2$\\
47~Tuc & $13.50\pm0.05$ & $13.62\pm0.08$ & $13.68\pm0.15^3$\\
\hline\noalign{\smallskip}
\noalign{$^1$Reid 1997; $^2$Chaboyer et al.~1997; $^3$Reid 1998}
\end{tabular}
\end{flushleft}
\end{table}

In Table~\ref{t:hd} we show the possible comparisons.  The error bars
for our ZAHB distances take into account the error on the
observational ZAHB location (as given in Paper I) and the small
contribution due to an uncertainty by $\pm$0.10 dex in the cluster
metallicity.  From Gratton et al.\ (1997) we adopted the binary
corrected distance moduli reported in column 8 of their Table 3, while
from Reid (1997) we have taken the distance moduli reported in column
7 of his Table 3 (derived by selecting the subsample of subdwarfs with
the most accurate parallax) for M5. In the case of M68, due to the
paucity of metal poor sub\-dwarfs considered by Reid (1997), we report
his result given in column 9 of the same table (all subdwarfs shifted
to a monometallicity sequence corresponding to the metallicity of
M68). Since Reid (1997) gives results for different reddening
estimates, we could select distances for those reddenings equal (or as
close as possible) to the one obtained from our isochrones.

For M68 and M5 our distances are in very good agreement with Gratton
et al.\ (1997), Reid (1997) and Chaboyer et al.\ (1997). The agreement
in the case of M5 (that is the cluster with the best available
photometry) is impressive.  One should note that the reddening and
cluster metallicity in these papers differ slightly from ours in some
cases. Both affect the distance determination by MS
fitting, while our distance modulus is independent of the reddening,
as long as the uncertainty is of the order of some hundredths of
magnitude. However, the differences in reddening are always $\le 0.01$
mag, and those in metallicity $< 0.10$ dex, such that
the results of the comparison are not modified appreciably.

A slightly different result is found in the case of 47~Tuc, for which
one notes less consistency between our distance modulus and
that by Gratton et al.\ (1997) and Reid (1998), although the 
distances are still compatible within the respective errors bars. Part
of the difference can be explained by the fact that in the quoted
papers solar-scaled isochrones were used
(see also the discussion at the end of Sect.~3.4).

Gratton et al.\ (1997) also provide the following empirical relation
between ZAHB brightness at the RR Lyrae instability strip and
metallicity, based on the distances they obtained by means of the
MS fitting technique applied to their sample of 9 clusters,

$$\rm M_{V}(ZAHB) = (0.22\pm0.09)\cdot([Fe/H]-1.5) + (0.49\pm0.04)\,\,\,\,(1)$$ 

\noindent
From our theoretical models (considering the case with $\Delta{Y}/\Delta{Z} = 3$) 
we derive, at $T_{\rm eff}$=3.85:

$$\rm M_{V}(ZAHB) = 0.17 \cdot([Fe/H]-1.5) + 0.52 \,\,\,\,(2)$$

\noindent
in very good agreement with the empirical relation (Eq.~1).

We can therefore conclude (as we did in Paper~II) that our models
predict distances completely confirmed by independent,
HIP\-PARCOS-based methods.
For completeness we add that our distance moduli as compared to
Paper~II changed due to two reasons: firstly, we use higher
metallicities, which lower the distance modulus; secondly, the new
calibration of the bolometric correction scale increases it again. The
net result is only a very small change.

\begin{figure}
\includegraphics[scale=0.40,draft=false]{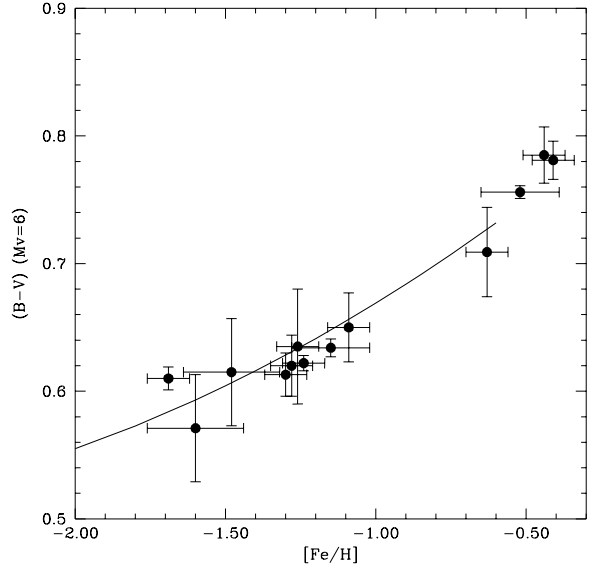}
\caption[]{ Run of the $(B-V)$ colour (at $M_{\rm v}$=6.0)
for unevolved MS stars as a function of [Fe/H].
The corresponding quantity from our theoretical isochrones is overimposed}
\protect\label{f:col}
\end{figure}

Before closing this section we briefly want to show the result of yet
another comparison between HIPPARCOS-based results and our theoretical
models. In this case we compare the $(B-V)$ colours of our
isochrones at $M_{\rm v}$=6.0 with the observed subdwarfs colour at
the same absolute magnitude, as derived by Gratton et al.\ (1997),
for a sample of 13 single and unevolved MS stars.
The result of this comparison is shown in Fig.~\ref{f:col}, where
empirical data (symbols) are compared with the corresponding
quantity from our theoretical isochrones (solid line).
The agreement between the colours of our isochrones and the
subdwarfs is very good at all metallcities.

\section{Discussion}

The original motivation for Paper~I had been to resolve the apparent
``age discrepancy'' between globular clusters and the expanding
universe. As we have demonstrated in all our papers, the oldest
globular clusters appear to be younger than 12 Gyr, possibly only as
old as 11 Gyr. Note that the neglect of diffusion even leads to a slight
overestimation of the ages by approximately 0.8 Gyr 
(Cassisi et al.\ 1998) for the vertical method 
(but the relative ages derived by means of the horizontal method are
unchanged). Such a cluster age is
completely consistent with the general range of recent
$H_0$-determinations (see Freedman 1997 for a review), which extends
from below 50 to 80 km/s/Mpc. Salaris \& Cassisi (1998)
used theoretical stellar models calculated with the same stellar
evolution program  
to predict the brightness of the tip of the RGB (their tip bolometric
luminosities are in complete agreement with our models), which
can be used as a primary distance indicator. Applying this to 
the galaxy NGC3379 in the Leo~I group, the distance to the Coma cluster
(obtained from the relative distance Coma-Leo I as derived from
different secondary distance indicators)
then gives a Hubble constant $H_0=60\pm11$ km/s/Mpc. Depending on
the cosmological model, the age of the universe is marginally or
completely consistent with our oldest clusters. Since an
open universe with $\Omega=0.3$ and $\Lambda=0$ appears to be preferred
(the probably most important single evidence coming from the
statistics of giant arcs produced by cluster of galaxies; see
Bartelmann et al.\ 1998), $H_0=60$ implies an age of $\approx$13 Gyr. 
Therefore distance indicator and cluster age result in a completely consistent
picture. 

\begin{figure}
\includegraphics[scale=0.40,draft=false]{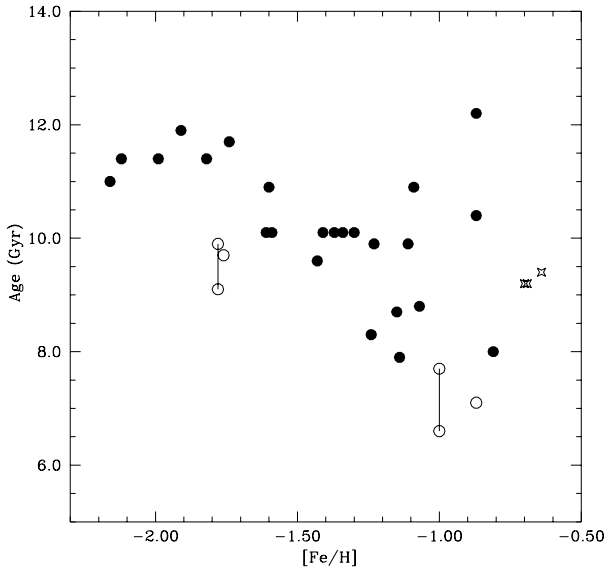}
\caption[]{The relation between age and metallicity for all clusters
dated in Papers I, II and the present one (Table~2). Filled circles
mark generic halo clusters, open circles clusters probably associated
with satellite galaxies or affected by tidal interaction with the
Magellanic Clouds. Crosses are the three disk clusters. For Rup106 and
Pal12 ages obtained under the assumptions of $\alpha$-element
enhanced or scaled-solar metals are connected by thin lines.}
\protect\label{f:afe}
\end{figure}

\begin{figure}
\includegraphics[scale=0.40,draft=false]{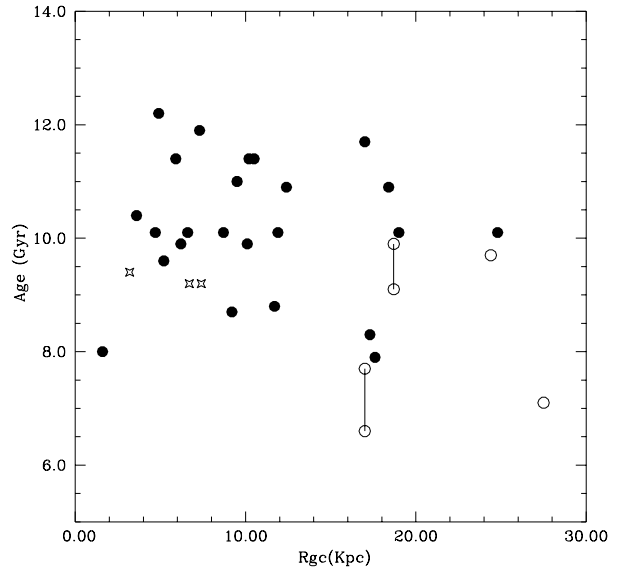}
\caption[]{ As in Fig.~\ref{f:afe}, but this time
the relation between age and galactocentric distances is shown.}
\protect\label{f:adi}
\end{figure}

The second purpose of our age determinations is to establish a
reliable age-metallicity relation for galactic globular clusters. From
Fig.~\ref{f:afe} the general scenario of Paper~II is confirmed and
extended: it appears that the more metal-poor clusters formed between
11 and 12 Gyr ago within a timespan of less than 1 Gyr. After that
halo clusters continued to be formed at higher metallicity for about 4
Gyr, with a tendency for more metal-rich clusters to form later (with
the exception of NGC6366, but see the discussion in Paper II about
this cluster).  The halo cluster creation continued even after the
disk had already been formed, as it is evident from the ages of the
oldest disk white dwarfs (Salaris et al.\ 1997; Leggett et al.\ 1997),
which are around 9 Gyr old. In the present paper, we found that the
three disk clusters investigated are coeval with these disk white dwarfs.
This is an important result, because previous age estimates (e.g.\
Chaboyer \& Kim) placed disk clusters at the same age as our oldest
halo clusters and let them appear to be considerably older than the
disk white dwarfs.

In addition, for
all metallicities there are clusters (e.g.\ Rup106 \& Arp2) that
appear to have been created at a later time, probably due to tidal
interactions with dwarf spheroidal galaxies (namely the Sagittarius
dSph) or the Magellanic Clouds. Such events are consistent with the
finding that cluster formation in the LMC apparently has been
triggered by close encounters with the Galaxy (Girardi et al.\ 1995;
Fujimoto \& Kumai 1997).

With respect to the age-galactocentric distance relation
(Fig.~\ref{f:adi}), our
conclusion of Paper~II is supported: within the innermost 10 kpc an
age spread of only $\approx 2.5$ Gyr exists, if one neglects the
exceptional clusters NGC6652 and NGC6366; the mean being at $\approx$11 Gyr. The
three disk clusters fit into this range at the lower boundary. At
larger distances, the age differences approach $\approx 5$ Gyr, with
the mean age being somewhat smaller. 
We agree with the conclusion of Sarajedini et
al.\ (1997) in that there appears to be a tendency that the innermost
parts of the galactic halo formed within a timespan shorter than
that for the outermost regions. However, we find the age spreads to
differ only by a factor of two, and given the small number of
clusters, it is not clear whether this is significant enough.
In particular, the larger spread for the outer halo clusters again depends on
a few clusters. One of these is Pal12, for which there are
indications (Brown et al.\ 1997) that it is not $\alpha$-enhanced. 
If true, its age would be raised by $\approx 1.0$ Gyr (Sect.~4.1),
making both the intermediate metal-rich group and the outer halo
clusters more homogeneous in age. Another cluster with uncertain
$\alpha$-enhancement is Rup106.

The disk clusters are different from the halo clusters with respect to
their composition. We have found strong evidence
that they have a solar-like helium content, while there is no
indication of this in the halo clusters of comparable
metallicity. Recalling that their metallicity is only half the solar
one, this implies a different chemical enrichment history for them
than for the solar neighbourhood.

We also showed that $\alpha$-element enhancement has to be
taken into account properly in all aspects of the theoretical
calculations for total metallicities $Z > Z_\odot/10$, in order to
obtain reliable isochrones.  
If the $\alpha$-elements are enhanced
in such a way that the condition [C+N+O+Ne/ Mg+Si+S+Ca+Fe]$\approx$0
is violated, a fact that is compatible with current available
observations, and the isochrone colours
have to be determined with high accuracy for all evolutionary stages,
then  the computation of  $\alpha$-enhanced isochrones 
is necessary even for the lowest metallicities.

Finally, distances predicted from our age determinations and ZAHB
models are in very good agreement with HIPPARCOS-based data,
demonstrating once more the reliability of our results.

\begin{acknowledgements}
We acknowledge stimulating and illuminating discussions with
R.~Peterson and R.D.~Cannon about observational details in general and
47~Tuc in particular. J\o rgen Christen\-sen-Dalsgaard helped to recover
Grundahl's thesis work. L.~Pulone generously made available a copy of
his paper prior to publication. The work of one of us (M.S.) was
carried out as part of the TMR programme (Marie Curie Research
Training Grants) financed by the EC.
\end{acknowledgements}

\end{document}